\pdfoutput=1
\documentclass[11pt,a4paper,english]{article}
\usepackage[T1]{fontenc}
\usepackage[latin1]{inputenc}
\usepackage{float}
\usepackage{amsmath}
\usepackage{graphicx}
\usepackage{setspace}
\usepackage{amssymb}

\makeatletter
\newcommand{\lyxaddress}[1]{
\par {\raggedright #1
\vspace{1.4em}
\noindent\par}
}

\usepackage{a4}

\usepackage{babel}
\makeatother
\begin{document}

\title{Special Relativity and possible Lorentz violations consistently coexist
in Aristotle space-time}

\author{Bernard Chaverondier}

\maketitle
\begin{center}Submitted to Foundation of Physics on 9 may 2008\par\end{center}

\lyxaddress{\begin{center}CNIM, ZI BREGAILLON, BP208, 83507 la Seyne sur mer
CEDEX, France\par\end{center}}

\begin{description}
\item [{Abstract:}] Some studies interpret quantum measurement as being
explicitly non local. Others assume the preferred frame hypothesis.
Unfortunately, these two classes of studies conflict with Minkowski
space-time geometry. On the contrary, in Aristotle space-time, Lorentz
invariance, interpreted as a physical property applying to all phenomena
actually satisfying this symmetry (as opposed to a geometrical constraint
applying to an assumed pre-existing Minkowski space-time) consistently
coexists with possible Lorentz violations. Moreover, as will be pointed
out, the geometrical framework provided by Aristotle space-time is
in fact necessary to derive the Lorentz transformations from physical
hypotheses.
\item [{Keywords:}] Special Relativity, preferred frame, Aristotle space-time,
quantum measurement.
\end{description}

\section{{\large Quantum Non-locality and quantum preferred frame}}

Percival proved realistic%
\footnote{Realistic interpretations \cite{Schlosshauer} assume quantum collapse
to be an objective (i.e. observer independent) physical process. Contrary
to the Everett Many Worlds Interpretation, severely criticized by
Neumaier \cite{MWI} and Bell \cite{Bell}, they conflict with Lorentz
invariance.%
} interpretations of quantum collapse to violate Lorentz invariance
in Bell-type experiments \cite{Percival}. Henceforth, as suggested
by Bell, non-locally correlated quantum events can be interpreted
as faster-than-light interactions \cite{Hemmick,Bell} complying with
the causality principle provided it rests on the absolute chronological
order associated with a quantum preferred frame \cite{Bellghost}.
Similarly, EPR experiments performed in Geneva by the Group of Applied
Physics \cite{Gisin1,Gisin2} have been analyzed according to the
not Lorentz invariant%
\footnote{Hence incompatible with Minkowski space-time.%
} preferred frame hypothesis.

\section{{\large Other reasons suggesting a preferred frame}}

Selleri argues that some superluminal effects strongly suggest the
need for a preferred frame and its associated preferred chronology
\cite{Selleri}. Moreover, the scalar theory of gravitation of Arminjon
\cite{Arminjontest,ArminjonWEP}, investigates a preferred frame gravitation
approach as a possible way to make quantum and gravitation theories
fit together. The preferred frame, formalized as a field of time-like
unit vectors, is also used in the context of preferred frame theories
of gravity by authors such as Will and Nordtvedt \cite{NordvedtextPPN}
Eling and Jacobson \cite{Elingtimefield}. Moreover, some attention
was devoted to the preferred frame hypothesis by Kostelecky as a consequence
of possible Lorentz violations in High Energy Physics \cite{Kostelecky1,KosteleckyCPTviol}.
Now, Aristotle space-time will provide us with a geometrical framework
authorizing the peacefull coexistence of the preferred frame hypothesis
with the ubiquitous Lorentz invariance. Let us now define Aristotle
space-time's symmetry group.

\section{{\large The Poincaré, Galilei and Aristotle groups}}

Thanks to Noether's theorem, energy and linear momentum, as well as
angular momentum conservation laws, arise from the invariance of the
Lagrangian of dynamical systems respectively with regard to the group
of space-time translations and the group $SO(3)$ of spatial rotations.
The semi-direct product group, arising from these two groups, is the
so-called restricted Aristotle group \cite{Souriau}. It will be denoted
$SA(4)$.

This seven parameters group is also the direct product group of the
Special Euclidean group $SE(1)$ (i.e. the temporal translations of
the 1D affine Euclidean space $E^{1}$) and the Special Euclidean
group $SE(3)$ (i.e. the direct spatial isometries of the 3D affine
Euclidean space $E^{3}$). So, $SA(4)$ is the intersection of the
restricted Galilei and Poincaré groups. Hence, neither does it contain
Galilean boosts, nor Lorentzian ones. Its relevance is the following:

\begin{itemize}
\item the invariance requirement of physical laws with regard to Galilean
boosts conflicts with interactions propagating at a finite speed independent
of the speed of their source, hence in particular with electromagnetism
\item the invariance requirement of physical laws with regard to Lorentzian
boosts conflicts with interactions propagating at infinite speed.
\end{itemize}
On the contrary, Aristotle group of symmetry complies with interactions
propagating at the speed of light as well as possible faster-than-light
interactions%
\footnote{Or possible instantaneous actions at a distance caused by quantum
measurements \cite{Valentini1,Valentini2,Bellghost}.%
}.

\section{{\large Aristotle space-time}}

\subsection{Definition and foliation of Aristotle space-time}

\begin{figure}[H]
\begin{singlespace}
\begin{centering}\includegraphics[scale=0.6]{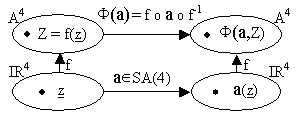}\par\end{centering}
\end{singlespace}

\caption{Aristotle space-time $A^{4}$}
\end{figure}
 Aristotle space-time, arising from the Aristotle group, embodies
only the Principle of Relativity with regard to space-time localization
and spatial orientation. It is defined as a set denoted $A^{4}$ equipped
with a bijection $f$ from $\mathbb{R}^{4}$ to $A^{4}$ providing
it with an action $\Phi$ of the numerical restricted Aristotle group
$SA(4)$%
\footnote{i.e. considered as a subgroup of $Gl_{4}(\mathbb{R})$.%
} defined as:\begin{equation}
\Phi:\begin{cases}
SA(4)\times A^{4} & \rightarrow A^{4}\\
(a,Z) & \mapsto\Phi_{a}(Z)=f\circ a\circ f^{-1}(Z)\end{cases}\label{eq:action}\end{equation}
From now on, we will identify $SA(4)$ with its representation acting
on $A^{4}$. So, for the sake of simplicity, $\Phi_{a}$, the image
of $a$ by action $\Phi$, will be identified with $a$ and $\Phi_{a}(Z)$
will be referred to as \emph{an} action $a$ of group $SA(4)$ on
$Z$.

Aristotle space-time is endowed with 2 preferred foliations

\begin{itemize}
\item a 1D foliation of which the 1D leaves of absolute rest are the orbits
of the time translation group, the invariant subgroup $SE(1)$ of
$SA(4)$.
\item a 3D foliation of which the 3D leaves of universal simultaneity are
the orbits of the direct isometries group, i.e. the invariant subgroup
$SE(3)$ of $SA(4)$.
\end{itemize}
As each one of these two foliations is a complete set of orbits of
an Aristotle invariant subgroup, this foliated structure is preserved
under Aristotle group actions. This foliation may be interpreted as
the preferred inertial frame of Bell's realistic interpretation of
quantum collapse and may also be helpful to account for possible Lorentz
violations \cite{Kostelecky1,KosteleckyCPTviol}.

\subsection{Spatial and temporal metrics of Aristotle space-time}

The quotient manifold of $A^{4}$ by its 1D foliation is diffeomorphic%
\footnote{The chosen manifold structure of $A^{4}$ is that induced by $f$,
i.e. diffeomorphisms of $A^{4}$ are bijections $F$ from $A^{4}$
to $A^{4}$ such that $f^{-1}\circ F\circ f$ are diffeomorphisms
of $\mathbb{R}^{4}$.%
} with the 3D leaves of universal simultaneity. This 3D manifold can
be equipped with an action of $SE(3)$ (the invariant subgroup of
spatial isometries). This provides it with the metric structure of
a 3D affine Euclidean space $E^{3}$. Similarly, the quotient manifold
of $A^{4}$ by its 3D foliation is diffeomorphic with the 1D leaves
of absolute rest and can be provided with the metric structure of
a 1D affine Euclidean space $E^{1}$. Hence, Aristotle space-time
can be identified as the Cartesian product $A^{4}=E^{1}\times E^{3}$.
It is naturally equipped with two Euclidean metrics which are invariant
with regard to the Aristotle group actions:

\begin{itemize}
\item a rank 1 temporal metric, which will be denoted $dT^{2}$
\item a rank 3 spatial metric, which will be denoted $dL^{2}$.
\end{itemize}

\subsection{Causal structure of Aristotle space-time}

The principle of relativity of motion, embodied in Minkowski space-time
geometry, forbids granting a privileged status to a preferred inertial
frame. Now, the chronological order between space-like separated events
depends on the rest inertial frame of the observer. This prevents
Minkowski space-time complying with the existence of causal links
spanning out of the light cone. On the contrary, Aristotle space-time
foliation into 3D leaves of universal simultaneity enables us to define
an objective chronology%
\footnote{That is to say independent of the motion of inertial observers.%
} between any pair of events. This gives rise to a causal structure
where possible faster-than-light interactions, comply with the principle
of causality prevailing in this space-time.

\section{{\large Aristotle charts and Aristotle bases}}

Aristotle space-time is associated with a family of preferred coordinate
systems, called Aristotle charts, preserving its foliated geometry
and its metrics.

\subsection{Aristotle charts}

In Aristotle space-time $A^{4}=E^{1}\times E^{3}$, any event $Z$
reads: $Z=(T,R)$

\begin{itemize}
\item $T\in E^{1}$ denotes the moment when event $Z$ occurs.
\item $R\in E^{3}$ denotes the localization where event $Z$ occurs.
\end{itemize}
Events $Z$ are localized in so-called Aristotle charts denoted $\mathcal{A}$
such that $Z=A(\underline{z})$, where $\underline{z}=(t,\underline{r})=(t,x,y,z)\in\mathbb{R}^{4}$
are the so-called coordinates of $Z$ in Aristotle chart $\mathcal{A}$.
By definition, Aristotle charts are such that:

\begin{itemize}
\item they preserve the foliation of Aristotle space-time into 1D lines
of absolute rest and 3D leaves of universal simultaneity. In particular,
two events belonging to a same simultaneity leave (i.e. occuring at
the same time $T$) have the same chronological coordinate $t$, i.e.
$\exists\mathcal{T}$: $\mathbb{R}\rightarrow E^{1}$ and $\exists\mathcal{R}$:
$\mathbb{R}^{3}\rightarrow E^{3}$ such that $\mathcal{A}(t,\underline{r})=(\mathcal{T}(t),\mathcal{R}(\underline{r}))$
\item the temporal metric be normalized, i.e. $dT^{2}=dt^{2}$
\item the spatial metric be Orthonormalized, i.e. $dL^{2}=dx^{2}+dy^{2}+dz^{2}$.
\end{itemize}
Besides, $O=E^{1}\times\{\mathcal{R}(\underline{0})\}$ will denote
the motionless observer resting at the spatial origin of Aristotle
chart $\mathcal{A}$ and $\{\mathcal{T}(0)\}\times E^{3}$ is the
3D leaf of universal simultaneity passing through origin event $E=\mathcal{A}(\underline{0})$
of chart $\mathcal{A}$.

\subsection{Aristotle bases}

Any Aristotle chart $\mathcal{A}$ is associated with a space-time
basis $\mathcal{V}=(\vec{t},\vec{x},\vec{y},\vec{z})$%
\footnote{From a differential geometry point of view, $\mathcal{V}=d\mathcal{A}$
\cite{Souriau}%
} of the vector space%
\footnote{$V^{1}\oplus V^{3}$ is the tangent space to $A^{4}=E^{1}\times E^{3}$%
} $V^{1}\oplus V^{3}$. Indeed, let us define

\begin{itemize}
\item $E=\mathcal{A}(\underline{0})$ the so-called origin event of chart
$\mathcal{A}$
\item Events $E_{t}=\mathcal{A}(1,\underline{0})$; $E_{x}=\mathcal{A}(0,1,0,0)$;
$E_{y}=\mathcal{A}(0,0,1,0)$; $E_{z}=\mathcal{A}(0,0,0,1)$
\item Unit vectors $\vec{t}=\overrightarrow{EE_{t}}$; $\vec{x}=\overrightarrow{EE_{x}}$;
$\vec{y}=\overrightarrow{EE_{y}}$; $\vec{z}=\overrightarrow{EE_{z}}$
\end{itemize}
$\vec{t}$ is a normalized vector of $V^{1}$ and $\mathcal{B}=(\vec{x},\vec{y},\vec{z})$
an Orthonormalized basis of $V^{3}$.

\subsection{Change of Aristotle charts}

Let $\mathcal{A}$ be an Aristotle chart. Let $\Phi$ denote the action
(\ref{eq:action}) of the restricted Aristotle group $SA(4)$ on $A^{4}$.
Any action $\Phi_{a}=\Phi(a)$ (denoted $a$ for the sake of simplicity),
of $a\in SA(4)$ on $A^{4}$ entails an Aristotle chart change\begin{equation}
(a,\mathcal{A})\rightarrow\mathcal{A}_{a}=a\circ\mathcal{A}=\mathcal{A}\circ\varphi_{a}\end{equation}

This definition ensures coordinates' covariance, i.e. the same system
will be located by the same coordinates whenever observer and observed
system both undergo the same chart change $\mathcal{A}\rightarrow\mathcal{A}_{a}$.
Coordinates $\underline{z}$ of the <<new event>> $Z=a(Z_{0})$
in the <<new chart>> $\mathcal{A}_{a}$ are the same as coordinates
of the <<old event>> $Z_{0}$ in the <<old chart>> $\mathcal{A}$.
$\varphi_{a}$ is the numerical expression of action $a$ in chart
$\mathcal{A}$.%
\begin{figure}[H]
\begin{singlespace}
\begin{centering}\includegraphics[scale=0.6]{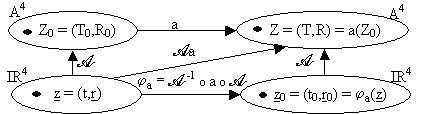}\par\end{centering}
\end{singlespace}

\caption{Change of Aristotle chart}
\end{figure}

\section{{\large Boosts and inertial charts in Aristotle space-time}}

So as to express Lorentz invariance of the phenomena that actually
satisfy this symmetry, we have to define Lorentzian boosts, inertial
charts and then to derive Lorentz transformations in Aristotle space-time
framework.

\subsection{Definition of boosts and pure boosts}

\subsubsection{Physical requirements applying to any boost\label{sub:boosts properties}}

We define boosts as diffeomorphisms of $A^{4}$ having the following
physical properties%
\footnote{They will be translated mathematically in sub-section \ref{sub:Mathematical-properties-of}%
}

1/ When applied to motionless observers, boosts set them in motion
with the same velocity $\vec{v}$ called velocity of the boost.

2/ The modification of durations and distances caused by the application
of a boost in the vicinity of a boosted event <<is the same>> whatever
this event.

3/ Freely moving observers%
\footnote{A freely moving observer is a line $\mathcal{D}$ with a direction
$D_{\vec{v}}=\{ t\vec{t}+t\vec{v}/t\in\mathbb{R}\}$, where $||\vec{v}||<c$
(the speed of light). $\vec{v}\in V^{3}$ is called the velocity of
this observer so that the rest lines of Aristotle space-time are observers
freely moving with a zero velocity.%
} keep on freely moving after the action of a boost.

\begin{singlespace}
4/ The covariance of boosts' observed effect is satisfied under any
change of Aristotle chart. Loosely speaking, a boost has the same
effect whatever the Aristotle chart where it is applied%
\footnote{This requirement expresses the homogeneity, the stationarity and the
isotropy of Aristotle space-time physical properties with regard to
boosts.%
}. Actually, Aristotle covariance of boosts will be assumed to hold
when, more generally, $a$ is any action of the complete Aristotle
group $A(4)$.%
\begin{figure}[H]
\begin{singlespace}
\begin{centering}\includegraphics[scale=0.6]{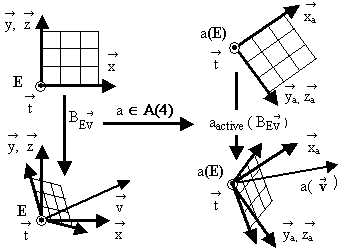}\par\end{centering}
\end{singlespace}

\caption{Covariance of boosts' effect under any Aristotle group action $a$}
\end{figure}

\end{singlespace}

5/ Symmetry of point of view between motionless and moving observers:
loosely speaking, we cancel the effect of a boost of velocity $\vec{v}$,
applied to Aristotle space-time $A^{4}$, by applying a boost of velocity
$-\vec{v}$%
\footnote{So, physical observers at rest in a boosted Aristotle chart (by a
boost of velocity $\vec{v}$), observing phenomena occurring in an
Aristotle chart $\mathcal{A}$, will observe the same effects as motionless
observers (hence at rest in $\mathcal{A}$) observing the same phenomena
occurring in an Aristotle chart boosted with the velocity $-\vec{v}$.
This assumption expresses the impossibility facing a steadily translating
observer if he tries to detect his absolute motion when using measurements
and phenomena that are Lorentz-covariant.%
}. 

6/ The maximal propagation speed measured by a motionless observer
is the same as that measured by an observer at rest in a boosted Aristotle
chart.

\subsubsection{Requirements applying specifically to pure boosts\label{sub:pure-boosts}}

So as to define the so-called pure boosts, i.e. boosts that are not
combined with Aristotle group actions, we ask for the following additional
properties:

7/ Any pure boost is endowed with at least one so-called origin event
$E$, invariant under the pure boost action. Thus, a pure boost, combined
with a spatial translation perpendicular to its velocity, is not anymore
a pure boost%
\footnote{Under a space-time translation, of which the translation vector is
included in the $(\vec{t},\vec{v})$ plane, the origin event $E$
of a pure boost shifts but the new boost is still a pure boost.%
}.

8/ Any pure boost is completely determined given its velocity and
an origin event. Together with property 4/ (the Aristotle covariance
of boosts) this makes it possible to eliminate pure boosts combined
with rotations.

9/ If $\{ B_{E\lambda\vec{v}}/\lambda\in\mathbb{R}\}$ is a family
of pure boosts having a same origin event $E$ and velocities proportional
to $\vec{v}$, $B_{E\lambda\vec{v}}\rightarrow i_{A^{4}}$ when $\lambda\rightarrow0$.
Together with the other properties, in particular the symmetry of
point of view, this will enable us to eliminate pure boosts combined
with $P$ or $T$ symmetries.

\subsection{Definition of inertial charts\label{sub:Inertial-frames}}

With any pure boost $B_{\vec{v}}$ of velocity $\vec{v}$, of origin
event $E$, and with any Aristotle chart $\mathcal{A}$ of same origin
event $E$, we associate a so-called inertial chart $\mathcal{A}_{\vec{v}}$
moving with the velocity $\vec{v}$. The chart $\mathcal{A}_{\vec{v}}$
is defined so as to ensure coordinates' covariance with regard to
boosts, i.e. if the <<old>> event $Z_{0}$ is localized by coordinates
$\underline{z}=(t,\underline{r})$ in the <<old>> chart $\mathcal{A}$,
then the <<new>> event $Z=B\vec{v}(Z_{0})$ is also localized by
these same coordinates in the <<new>> chart $\mathcal{A}_{\vec{v}}$.
So: %
\begin{figure}[H]
\begin{singlespace}
\begin{centering}\includegraphics[scale=0.6]{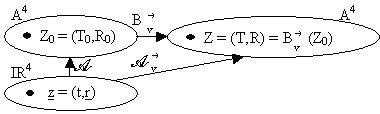}\par\end{centering}
\end{singlespace}

\caption{Definition of an inertial chart $\mathcal{A}_{\vec{v}}$}
\end{figure}
 $Z_{0}=\mathcal{A}(\underline{z})\Rightarrow B_{\vec{v}}(Z_{0})=Z=\mathcal{A}_{\vec{v}}(\underline{z})$.
So that \begin{equation}
\mathcal{A}_{\vec{v}}=B_{\vec{v}}\circ\mathcal{A}\end{equation}
Moreover $\mathcal{A}_{\vec{v}}(\underline{0})=B_{\vec{v}}(\mathcal{A}(\underline{0}))=B_{\vec{v}}(E)=E$,
so that $\mathcal{A}_{\vec{v}}$ has same space-time origin $E$ as
$\mathcal{A}$.

The expression $b_{\underline{v}}$ of boost $B_{\vec{v}}$ in Aristotle
chart $\mathcal{A}$ will be defined as follows:%
\begin{figure}[H]
\begin{singlespace}
\begin{centering}\includegraphics[scale=0.58]{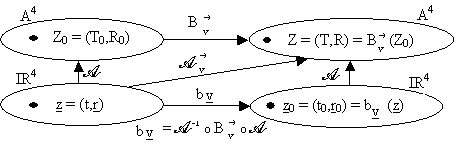}\par\end{centering}
\end{singlespace}

\caption{Expression of a boost $B_{\vec{v}}$ in an Aristotle chart $\mathcal{A}$}
\end{figure}

$\underline{z}_{0}=(t_{0},\underline{r}_{0})=b_{\underline{v}}(\underline{z})$
are the coordinates of event $Z=B_{\vec{v}}(Z_{0})$ in the <<old
chart>> $\mathcal{A}$ wheras $\underline{z}$ are its coordinates
in the <<new chart>> $\mathcal{A}_{\vec{v}}$. The covariance of
coordinates $\underline{z}$ with regard to boost $B_{\vec{v}}$ means
that the passive transformation $\mathcal{A}\rightarrow\mathcal{A}_{\vec{v}}$
(causing the change of coordinates $\underline{z}_{0}\rightarrow\underline{z}$)
caused by boost $B_{\vec{v}}$ when applied to the observer only (i.e.
to the observation frame $\mathcal{A}$ and not to the observed system)
cancels the active transformation $Z_{0}\rightarrow Z=B_{\vec{v}}(Z_{0})$
(causing the change of coordinates $\underline{z}\rightarrow\underline{z}_{0}$),
i.e. the action of this same boost $B_{\vec{v}}$ when applied to
the observed system only. Besides, we notice that, thanks to the choice
of a chart $\mathcal{A}$ that has same origin event $E$ as boost
$B_{E\vec{v}}$:\[
b_{\underline{v}}(\underline{0})=\mathcal{A}^{-1}\circ B_{E\vec{v}}\circ\mathcal{A}(\underline{0})=\mathcal{A}^{-1}\circ B_{E\vec{v}}(E)=\mathcal{A}^{-1}(E)=\underline{0}\]

\subsection{Mathematical properties of pure boosts\label{sub:Mathematical-properties-of}}

Let us now translate mathematically the physical requirements of sub-section~\ref{sub:boosts properties}

1/ Motionless observers are set in motion with the velocity $\vec{v}$
of the boost.\\
If $\mathcal{M}=E^{1}\times\{ M\}$ (where $M\in E^{3}$) is a motionless
observer and $B_{\vec{v}}$ is a boost of velocity $\vec{v}$, then
$B_{\vec{v}}(\mathcal{M})$ is a line $\mathcal{D}$ of direction
$D_{\vec{v}}=\{ t\vec{t}+t\vec{v}/t\in\mathbb{R}\}$

2/ The effect of boost $B$ in the vicinity of event $B(Z)$ does
not depend on event $Z$. That is to say, $\forall\, Z_{1},Z'_{1},Z_{2},Z'_{2}$
such that $\overrightarrow{Z_{1}Z'_{1}}=\overrightarrow{Z_{2}Z'_{2}}$
and for any boost $B$: $\overrightarrow{B(Z_{1})B(Z'_{1})}=\overrightarrow{B(Z_{2})B(Z'_{2})}$%
\begin{figure}[H]
\begin{singlespace}
\begin{centering}\includegraphics[scale=0.56]{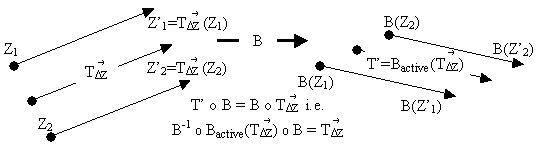}\par\end{centering}
\end{singlespace}

\begin{singlespace}

\caption{Effect of a boost on a translation}\end{singlespace}

\end{figure}

\begin{itemize}
\item Let $T=T_{\overrightarrow{\Delta Z}}$ be the translation of vector
$\overrightarrow{\Delta Z}=\overrightarrow{Z_{1}Z'_{1}}$\\
so that $Z'_{1}=T(Z_{1})$ and $Z'_{2}=T(Z_{2})$
\item Let $T'=T_{\overrightarrow{\Delta Z'}}$ be the translation of vector
$\overrightarrow{\Delta Z'}=\overrightarrow{B(Z_{1})B(Z'_{1})}$\\
so that $B(Z'_{1})=T'(B(Z_{1}))$ and $B(Z'_{2})=T'(B(Z_{2}))$
\end{itemize}
$\forall\, Z\in A^{4}$: $B(T(Z))=T'(B(Z))$, i.e. $B\circ T=T'\circ B$
and it is easy to establish that: $dB$ is constant (i.e. $B$ is
affine) and $B\circ T_{\overrightarrow{\Delta Z}}\circ B^{-1}=T_{dB(\overrightarrow{\Delta Z})}$

It is worth noticing that the above equation may also be written:\begin{equation}
B_{passive}\circ B_{active}(T_{\overrightarrow{\Delta Z}})=T_{\overrightarrow{\Delta Z}}\end{equation}

\begin{itemize}
\item the active transformation, $B_{active}(T_{\overrightarrow{\Delta Z}})=T_{dB(\overrightarrow{\Delta Z})}$,
of a space-time translation $T_{\overrightarrow{\Delta Z}}$ is a
change of the observed space-time translation effect (applied to a
given system) when the space-time translation vector $\overrightarrow{\Delta Z}$
as well as the observed system, both undergo the same boost $B$. 
\item The passive transformation, $B_{passive}(T')=B^{-1}\circ T'\circ B$,
of a space-time translation $T'$ is a change of the observed translation
effect when only the observer (i.e. the observation chart) undergoes
boost $B$. 
\end{itemize}
So, requirement 2/ amounts to the covariance of space-time translations
observed effects under any boost $B$ (ie the invariance of these
observed effects when the applied space-time translation, the observed
system as well as the observer all undergo the same boost $B$). This
proves requirement 2/ to express the principle of relativity of motion
with regard to translation observed effects. As seen above, this causes
boosts to be affine transformations.

Moreover, the expression $b$ of a pure boost $B$, in a chart $\mathcal{A}$
that has the same origin event $E$ as boost $B$, satisfies $b(\underline{0})=\underline{0}$%
\footnote{As concluded in sub-section \ref{sub:Inertial-frames}.%
}. Hence, $b$ is linear.

3/ Any freely moving observer keeps freely moving when boosted: as
boosts are affine transformations, they transform affine lines into
affine lines of Aristotle space-time so that physical requirement
3/ of sub-section \ref{sub:boosts properties} is satisfied. Actually,
there is an equivalence between the requirement 2/ and the requirement
3/ that lines of $A^{4}$ be transformed into lines of $A^{4}$ (we
may have preferred to derive 2/ from 3/ instead of the other way around).

4/ Covariance of boosts observed effects under any change of Aristotle
chart:

\begin{itemize}
\item Let us define the active transformation of a pure boost $B=B_{E\vec{v}}$
of velocity $\vec{v}$ and origin event $E$ under an action $a\in SA(4)$
as a pure boost of origin event $a(E)$ and velocity $da(\vec{v})$,
i.e.\begin{equation}
a_{active}(B_{E\vec{v}})=B_{a(E)da(\vec{v})}\end{equation}
 Physically, this transformation represents an action $a$ both on
the applied boost and on the observed system%
\footnote{But not on the observer.%
}.
\item Let us now define a passive transformation of any boost $B'$ as:\begin{equation}
a_{passive}(B')=a^{-1}\circ B'\circ a\end{equation}
 Physically, this transformation represents an action $a$ on the
observer only, i.e. a change $\mathcal{A}\rightarrow\mathcal{A}_{a}$
of Aristotle chart of observation.
\end{itemize}
The assumed covariance of observed boost effects when observer, observed
system and applied boost all undergo the same Aristotle chart change
$a$ reads%
\footnote{To exemplify the physical meaning of Lorenzian boosts' covariance
property (with regard to any action $a$ of the Aristotle group),
let us consider the special case of the covariance with regard to
spatial rotations. So, let us consider, for instance, a strain tensor
field induced in an \emph{isotropic} 3D medium submitted to an homogeneous
(but anisotropic) stress tensor field. If we rotate \emph{both} the
observer and the applied stress tensor field, then, the passive transformation
(the rotation of the observer) cancels the active transformation (the
strain field modification induced by the rotation of the applied stress
tensor field). Because of this 3D medium isotropy, the rotated observer
will observe the same effect as if neither himself, nor the stress
tensor field had been rotated. It wouldn't be the case if this medium
were anisotropic. Similarly, we demand the invariance of space-time
deformations under any Lorentzian boost, when the boost undergoes
the combination of an active and a passive action of any Aristotle
group action $a$. This amounts to require space-time behaving as
an homogeneous, isotropic and stationary medium.%
}: \begin{equation}
a_{passive}\circ a_{active}(B)=B\end{equation}
i.e. $a^{-1}\circ a_{active}(B)\circ a=B$, so that \begin{equation}
a\circ B_{E\vec{v}}\circ a^{-1}=B_{a(E)da(\vec{v})}\end{equation}

Now, hypothesis 4/ of sub-section \ref{sub:boosts properties} requires
that the above condition must hold for any action $a$ of the complete
Aristotle group $A(4)$.

\section{{\large Derivation of Lorentz transformations}}

A rigorous derivation of Lorentz transformations from physical hypotheses
(cf sub-section~\ref{sub:boosts properties}) needs using Aristotle
space-time and its symmetries (cf sub-section~\ref{sub:Covariance1},
\ref{sub:Covariance2}) with regard to boosts effects.

Let us consider a boost, denoted $B_{\vec{v}}$, of velocity $\vec{v}$
and origin event $E$.\\
Let us consider an Aristotle chart $\mathcal{A}$ of origin event
$\mathcal{A}(\underline{0})=E$ and spatial origin $O$, having its
vector $\vec{x}$ in the same direction as velocity $\vec{v}$ (i.e.
$\vec{v}=v\vec{x}$).

\subsection{Covariance of a boost under a 180° rotation around $O\vec{x}$\label{sub:Covariance1}}

As a 180° rotation $R_{\pi\vec{x}}$ around $O\vec{x}$ neither changes
$E$ nor changes $\vec{v}$, $dR_{\pi\vec{x}}(\vec{v})=\vec{v}$ and\begin{equation}
R_{\pi\vec{x}}\circ B_{\vec{v}}\circ R_{\pi\vec{x}}=B_{\vec{v}}\end{equation}
 So that, in $\mathbb{R}^{4}$: $r_{\pi\underline{x}}\cdot b_{\underline{v}}\cdot r_{\pi\underline{x}}=b_{\underline{v}}$.
Now, in chart $\mathcal{A}$, matrix $r_{\pi\underline{x}}$ of $R_{\pi\vec{x}}$
reads:

\begin{center}\begin{equation}
r_{\pi\underline{x}}=\left\{ \begin{array}{cccc}
1 &  &  & 0\\
 & 1\\
 &  & -1\\
0 &  &  & -1\end{array}\right\} \end{equation}
\par\end{center}

The right multiplication of matrix $b_{\underline{v}}$ (of boost
$B_{\vec{v}}$) by matrix $r_{\pi\underline{x}}$ reverses the signs
of columns $y$ and $z$ of $b_{\underline{v}}$. The left multiplication
of matrix $b_{\underline{v}}$ by matrix $r_{\pi\underline{x}}$ reverses
the signs of lines $y$ and $z$ of $b_{\underline{v}}$. Consequently,
any off-diagonal $y$ and $z$ term of matrix $b_{\underline{v}}$
vanishes except the $yz$ terms.

\subsection{Covariance under a 90° rotation around $O\vec{x}$ axis and under
the $\Pi_{\vec{y}}=P\: R_{\pi\vec{y}}$ plane symmetry\label{sub:Covariance2}}

Similarly, we get $byy=bzz$ and $byz=bzy=0$, so that we have:

\begin{center}$\exists\, a,a',b',b"$ and $e\in\mathbb{R}$ such that:\begin{equation}
\left\{ \begin{array}{c}
t_{0}=at+a'x\\
x_{0}=b't+b''x\\
\begin{array}{cc}
y_{0}=ey\: & z_{0}=ez\end{array}\end{array}\right.\end{equation}
\par\end{center}

\subsection{Symmetry between motionless and moving observers}

\begin{itemize}
\item Applying boost $B_{-\vec{v}}$ erases boost $B_{\vec{v}}$, i.e. $B_{-\vec{v}}=B_{\vec{v}}^{-1}$,
\item The maximum propagation speed is covariant with regard to any Aristotle
group action, hence it is isotropic. Moreover, as far as Lorentz invariance
is satisfied, it has the same norm $c$ in $\mathcal{A}$ as in $\mathcal{A}\vec{v}$.
\end{itemize}
For convenience, let us introduce speed $c$ in the previously stated
equations:

\begin{center}$\exists\, a,a',b,b'$ and $e\in\mathbb{R}$ such that:
\begin{equation}
\left\{ \begin{array}{c}
ct_{0}=act+bx\\
x_{0}=b'(ct)+a'x\\
\begin{array}{cc}
y_{0}=ey\: & z_{0}=ez\end{array}\end{array}\right.\end{equation}
\par\end{center}

As $b_{\underline{v}}^{-1}=b_{-\underline{v}}$ and $b_{-\underline{v}}=\pi_{\underline{x}}\cdot b_{\underline{v}}\cdot\pi_{\underline{x}}$
(where $\pi_{\underline{x}}$ denotes the sign reversal of $\underline{x}$
):

\begin{center}$e^{-1}=e$ and \begin{equation}
[1/(aa'-bb')]\left\{ \begin{array}{cc}
a' & -b\\
-b' & a\end{array}\right\} =\left\{ \begin{array}{cc}
a & -b\\
-b' & a'\end{array}\right\} \end{equation}
\par\end{center}

Consequently $aa'-bb'=1$, $a=a'$ and $e^{2}=1$ so that $e=\pm1$.
Actually $e=1$. Indeed, according to requirement 9/ of sub-section
\ref{sub:pure-boosts}, pure boost $b_{\underline{v}}$ is assumed
to tend to the identity of $\mathbb{R}^{4}$ when $\vec{v}$ tends
to $\vec{0}$. Now, as the origin of $\mathcal{A}_{\vec{v}}$ (located
at $x=y=z=0$) moves with the velocity $\vec{v}=v\vec{x}$ we have
$x_{0}=vt_{0}$. As $x=y=z=0$ we have: $ct_{0}=a(ct)$ and $x_{0}=b'(ct)$.
Hence $a(vt_{0})=ax_{0}=ab'(ct)=b'(ct_{0})$, so that\begin{equation}
b'=av/c\end{equation}

Now, let us express the covariance of the relative speed $c$ of light:\\
$x_{0}=ct_{0}\Rightarrow x=ct$ so that $x_{0}=b'(ct)+a'x=ct_{0}=a(ct)+bx\Rightarrow$
\begin{equation}
b'+a'=a+b\end{equation}
As $a=a'$ we get $b=b'$. Hence $aa'-bb'=1$ becomes $a^{2}-b'^{2}=1$\\
As $b'=av/c$, this provides $a^{2}-(av/c)^{2}=1$ so that $a=\pm1/(1-v^{2}/c^{2})^{1/2}$\\
Now, we have excluded time reversal. Indeed, $b_{\underline{v}}$
is assumed to tend to the identity of $\mathbb{R}^{4}$ when $\underline{v}$
tends to $\underline{0}$ so that\begin{equation}
a=1/(1-v^{2}/c^{2})^{1/2}\end{equation}
 Finally, we get the Lorentz transformations:

\begin{center}\begin{equation}
\left\{ \begin{array}{c}
ct_{0}=(ct+vx/c)/(1-v^{2}/c^{2})^{1/2}\\
x_{0}=(vt+x)/(1-v^{2}/c^{2})^{1/2}\\
\begin{array}{cc}
y_{0}=y & \: z_{0}=z\end{array}\end{array}\right.\end{equation}
\par\end{center}

\section{{\large Conclusion}}

The present article exhibits Aristotle spacetime foliated structure,
its causal structure and the peaceful coexistence, in this arena,
of the phenomena actually satisfying Lorentz invariance with possible
Lorentz violations. It provides a geometrical framework where realistic,
hence explicitly non local interpretations of quantum collapse, comply
with the principle of causality and suggests the possibility of interpreting
Lorentz invariance as a thermodynamical statistical emergence. Last
but not least, Aristotle spacetime geometry modelizes the energy,
linear and angular momentum conservation laws. This first step is
in fact needed to derive rigorously the Lorentz transformations from
the observed relativity of motion.

\end{document}